\documentclass[11pt]{article}
\usepackage{times,graphics}

\usepackage[margin=0.8in]{geometry}

\title{Kernel Matrix Regression}

\author{
Yoshihiro Yamanishi\thanks{} \\
Bioinformatics Center\\
Institute for Chemical Research\\
Kyoto University\\
Gokasho, Uji, Kyoto 611-0011, Japan \\
\texttt{yoshi@kuicr.kyoto-u.ac.jp} \\
\and
Jean-Philippe Vert \\
Centre for Computational Biology\\
Ecole des Mines de Paris\\
35 rue Saint-Honore\\
 77305 Fontainebleau cedex, France\\
\texttt{Jean-Philippe.Vert@mines.org} \\
}

\newcommand{\xb}{{\bf x}}
\newcommand{\yb}{{\bf y}}
\newcommand{\ub}{{\bf u}}
\newcommand{\vb}{{\bf v}}
\newcommand{\wb}{{\bf w}}
\newcommand{\RR}{{\bf R}}

\newcommand {\alphab}{ {\mbox{\boldmath $\alpha$}} }
\newcommand {\betab}{ {\mbox{\boldmath $\beta$}} }

\begin{document}

\maketitle

\begin{abstract}
We address the problem of filling missing entries in a kernel Gram matrix, given a related full Gram matrix.
We attack this problem from the viewpoint of regression, assuming that the two kernel matrices can be considered as explanatory variables and response variables, respectively.
We propose a variant of the regression model based on the underlying features in the reproducing kernel Hilbert space by modifying the idea of kernel canonical correlation analysis, and we estimate the missing entries by fitting this model to the existing samples.
We obtain promising experimental results on gene network inference and protein 3D structure prediction from genomic datasets.
We also discuss the relationship with the {\it em}-algorithm based on information geometry.
\end{abstract}


\section{Introduction}
\label{sec:intro}

Kernel methods such as support vector machines (SVM) enable the use of powerful statistical analysis for various datasets as soon as kernel matrices for the dataset are available [1].
When a dataset contains $N$ objects, the $N$ objects are represented as an $N \times N$ positive semidefinite matrix whose elements can be thought of as objects-object similarities.
An advantage of kernel methods is that they can be applied not only to real-valued data but also to complex structured objects such as strings, trees and graphs
[2].
The kernel matrix not only plays an important role as the input to kernel methods, but also provides important information regarding the similarity between objects.

In this paper we consider the problem of estimating missing entries in a kernel matrix.
More precisely we assume that two datasets describing the same objects are available; however, although all data are available for the first dataset, only part of the second dataset is available.
In this case, a full kernel Gram matrix can be obtained for the first dataset, while only a partial kernel Gram matrix is obtained for the second dataset, that is, a matrix with missing entries. If we are more interested in the second dataset rather than in the first dataset, it is natural to think of estimating the missing part of the second kernel matrix, e.g., by looking for correlations between the two kernels.

This problem arises commonly in  applications such as bioinformatics, where informative data for a given classification task is expensive to produce while less informative data are easily available. 
As an example, the DNA sequences of all proteins of a given organism are easily obtained from the sequenced genome, while the 3D structures of most proteins are still unknown and difficult to obtain. 
In this case, we want to know the 3D structure information for the proteins whose structure have not been determined. 
As another example, high-throughput genome-wide data (e.g., gene expression data) are available for the full genes of an organism, while the metabolic network information is known only for a limited number of genes.
In this case, we want to predict the unknown part of gene network using the fully available genome-wide data. 
A variety of methods have been proposed in order to tackle such problems related to kernel matrix completion.
Examples include the use of kernel canonical correlation analysis (kernel CCA) for gene classification [3] and gene network inference [4], and the use of the {\it em}-algorithm based on information geometry for protein 3D structure prediction [5].

In this paper we attack the kernel matrix completion problem from the viewpoint of regression, considering the two kernel matrices represent inner products between explanatory and response variables, respectively.
We propose a variant of the regression model based on the underlying features in the reproducing kernel Hilbert space by modifying the idea of kernel CCA, and we estimate the missing entries by fitting this model to the existing samples.
In the experiment, we show promising results on the prediction of missing edges in a gene network from genome-wide data and on the prediction of protein 3D structure information from their sequences.
We also discuss the mathematical and numerical relationships between the proposed methods and the {\it em}-algorithm based on information geometry.

\section{Formalism of the question}
\label{sec:formalism}

Suppose that we have an explanatory random variable $\xb \in \RR^d$ and a response random variable $\yb \in \RR^l$.
Let us now consider the situation where the data is available for all the $N$ objects for the explanatory variable $\xb$, while the data is available for the first $n$ objects and unavailable for the remaining ($N-n$) objects for the response variable $\yb$.
We refer to the first $n$ objects as {\it training set}, and we refer to the remaining $N-n$ objects as {\it test set} below.

Let $k$ and $g$ be symmetric positive definite kernels defined on $\RR^d$ and $\RR^l$, respectively. When we compute the kernel matrix for the explanatory variable $\xb$, we obtain an $N \times N$ kernel matrix $K$, where $(K)_{ij}=k(\xb_i,\xb_j) \,\, (1 \le i,j \le N)$, $\xb_i$ belongs to a set $\mathcal{X}$ and $N$ is the number of all objects.
On the other hand, when we compute the kernel matrix for the response variable $\yb$, we obtain an $N \times N$ kernel matrix $G$, where $(G)_{ij}=g(\yb_i,\yb_j) \,\, (1 \le i,j \le n)$, $\yb_i$ belongs to a set $\mathcal{Y}$, and $n$ is the number of available objects ($n < N$).
Note that $G$ contains in fact missing values for all entries $(G)_{ij}$ with $\max(i,j)>n$. 
We want to estimate the missing part of $G$ using full Gram matrix $K$, taking into account a form of correlation between the two kernels.

In this study we express each kernel matrix by splitting the matrix into four parts.
We denote by $K_{tt}$ (resp. $G_{tt}$) the $n \times n$ kernel matrix for the {\it training set} versus itself, $K_{pt}$ (resp. $G_{pt}$) the $(N-n) \times n$ kernel matrix for the {\it test set} versus the {\it training set}, and $K_{pp}$ (resp. $G_{pp}$) the $(N-n) \times (N-n)$ kernel matrix for the {\it test set} versus itself:
\begin{equation}
K =
\left(
  \begin{array}{cc}
  K_{tt} & K_{pt}^\top \\
  K_{pt} & K_{pp}
  \end{array}
\right)
, \quad
G =
\left(
  \begin{array}{cc}
  G_{tt} & G_{pt}^\top \\
  G_{pt} & G_{pp}
  \end{array}
\right)
\end{equation}
Note that $K_{pt}$ and $K_{pp}$ are known, while $G_{pt}$ and $G_{pp}$ are unknown. The goal is to predict  $G_{pt}$ and $G_{pp}$ from $K$ and $G_{tt}$.


\section{Methods}

In this section we describe four approaches that can be used for the problem of kernel matrix completion: the direct approach (Section \ref{sec:direct}) is a baseline straightforward approach, the kernel CCA approach (Section \ref{sec:super_CCA}) has been proposed in previous work [4], while the kernel matrix regression (Section \ref{sec:KMR}) and penalized kernel matrix regression (Section \ref{sec:PKMR}) are new.

\subsection{Direct approach}
\label{sec:direct}

A straightforward approach is to directly plug the entries of the kernel matrix $K$ for the explanatory variable $\xb$ into the missing entries of the kernel matrix $G$ for the response variable $\yb$, that is, to choose $G_{pt} = K_{pt}$ and $G_{pp} = K_{pp}$.
We refer to this approach as the direct approach.

\subsection{Kernel CCA (kCCA)}
\label{sec:super_CCA}

The use of kernel canonical correlation analysis (kCCA) has been proposed to estimate the unknown part of the metabolic network form genomic data [4]. We make a brief review of this approach in this subsection.

This approach amounts to searching low-dimensional feature spaces derived from both kernels that are maximally correlated during the training phase. The reconstruction of missing entries in $G$ is then obtained by projecting the corresponding points onto the feature space for the kernel $K$, and computing their inner product in this feature space as an approximation of the kernel $G$. More formally, let us write a feature $u:\RR^d\rightarrow \RR$ for the explanatory variable $\xb$, and a feature $v:\RR^l\rightarrow\RR$ for the response variable $\yb$ in the reproducing kernel Hilbert spaces as follows: 
\begin{equation}
u(\xb) = \sum_{j=1}^n k(\xb, \xb_j) \alpha_j,
\quad 
v(\yb) = \sum_{j=1}^n g(\yb, \yb_j) \beta_j,
\end{equation}
and use the notation $\alphab=(\alpha_1,\alpha_2,\cdots,\alpha_n)^\top$ and $\betab=(\beta_1,\beta_2,\cdots,\beta_n)^\top$.
The objective is to find features $u$ and $v$ that are as correlated as possible, that is,  which maximize the following correlation
\begin{equation}
\rho = Corr(u,v) = \frac{Cov(u,v)}{Var(u)^{1/2} Var(v)^{1/2}},
\end{equation}
where $Cov(u,v)=E(uv)-E(u)E(v)$ and $Var(u)=E(u^2)-E(u)^2$.
For theoretical and practical reasons [6] it is better to compute features $u$ and $v$ which maximize the following penalized canonical correlation:
\begin{equation}
\rho = 
\frac{\alphab^\top K_{tt} G_{tt} \betab}
{ \sqrt{\alphab^\top(K_{tt} + \lambda_x I )^2 \alphab}
  \sqrt{\betab^\top(G_{tt} + \lambda_y I )^2 \betab} },
\end{equation}
where $I$ is an identity matrix, and $\lambda_x$ and $\lambda_y$ are positive regularization parameters, and the matrices $K_{tt}$ and $G_{tt}$ are assumed to be centered.
When $m$ different features $u^{(j)}$ and $v^{(j)}$ associated the $m$ largest canonical correlations $\rho^{(j)}, j=1,2,\cdots,m$ are obtained, they can be merged into feature vectors $\ub(\xb)$ and $\vb(\yb)$ as 
$\ub(\xb) = (u^{(1)}(\xb),\cdots,u^{(m)}(\xb))^\top$ and 
$\vb(\yb) = (v^{(1)}(\yb),\cdots,v^{(m)}(\yb))^\top$.
The missing entries in $G$ are then estimated as 
$\hat{g}(\yb, \yb^\prime) = \ub(\xb)^\top \ub(\xb^\prime)$.

\subsection{Kernel matrix regression (KMR)}
\label{sec:KMR}

An apparent drawback of the kCCA approach is that the objective function of kCCA is different from that of correctly predicting the values of the kernel $G$. In particular, by computing  features $\vb$ for the response variable $\yb$, the notion of similarity between response variable $\yb$ is changed. In the problem of kernel matrix completion, we do not want to change the similarity space for the response variable $\yb$.
We want instead to change the object-object similarity space only for the explanatory variable $\xb$ to make it fit the the object-object similarity space for the response variable $\yb$. In this section we propose a variant of the regression model based on the underlying features in the reproducing kernel Hilbert space by modifying the idea of kernel CCA.

The ordinary regression model between an explanatory variable $\xb \in \RR^d$ and a response variable $y \in \RR$ can be formulated as follows:
\begin{equation}
y = f(\xb) + \epsilon, 
\end{equation}  
where $f: \RR^d \to \RR$ and $\epsilon$ is a noise term.
By analogy we propose to regard $(\xb, \xb^\prime) \in \RR^d \times \RR^d$ as an explanatory variable and $g(\yb,\yb^\prime) \in \RR$ as a response variable in our context.
Assuming the underlying feature $\ub(\xb) \in \RR^m$ in the reproducing kernel Hilbert space, we formulate a variant of the regression model as follows:
\begin{equation}
g(\yb, \yb^\prime) = f(\xb, \xb^\prime) + \epsilon 
                   = \ub(\xb)^\top \ub(\xb^\prime) + \epsilon,
\end{equation}  
where $f: \RR^d \times \RR^d \to \RR$.
We refer to this model as kernel matrix regression model. We note that imposing $f$ to be of the form $ f(\xb, \xb^\prime) =  \ub(\xb)^\top \ub(\xb^\prime)$ for some feature $\ub:\RR^d\rightarrow \RR^m$ ensures that the regression function is positive definite and the number of dimension $m$ of the feature $\ub$ is allowed to be infinite.

Following a classical approach in kernel methods, we consider features  in the reproducing kernel Hilbert space of the kernel $K$ that possess an expansion of the form: 
\begin{equation}\label{eq:feature}
u(\xb) = \sum_{j=1}^n k(\xb, \xb_j) w_j,
\end{equation} 
where $\wb=(w_1,w_2,\cdots,w_n)^\top$ is a weight vector and $n$ is the number of objects in the training set.
When $m$ different features are considered, we express them by a feature vector $\ub$ as $\ub(\xb) = (u^{(1)}(\xb),u^{(2)}(\xb),\cdots,u^{(m)}(\xb))^\top$.

In order to represent the set of features for all the objects, we define feature score matrices $U_t(\xb)=[\ub(\xb_1),\ub(\xb_2),\cdots,\ub(\xb_n)]^\top$ for the training set and $U_p(\xb)=[\ub(\xb_{n+1}),\ub(\xb_{n+2}),\cdots,\ub(\xb_{N})]^\top$ for the test set.

%

In the matrix form, we can actually compute the feature score matrices as $U_t = K_{tt} W$ for the training set and $U_p = K_{pt} W$ for the test set, 
where $W=[\wb^{(1)},\wb^{(2)},\cdots,\wb^{(m)}]$.

%

The inner products of the feature vectors between two objects are 
$q(\xb, \xb^\prime) = \ub(\xb)^\top \ub(\xb^\prime)$.
To represent all the object-object similarities in the feature space, we define the similarity matrix $Q$ as
$$
(Q)_{ij} = q(\xb_i, \xb_j) = \ub(\xb_i)^\top \ub(\xb_j), \quad 1 \le i,j \le N.
$$
Splitting the matrix $Q$ into several parts according to the training set, test set and their interaction, we can compute them as follows:

{\it Training set {\rm versus} Training set:}
\begin{equation}
Q_{tt} = U_t U_t^\top = K_{tt} W W^\top K_{tt}^\top\;,
\end{equation}

{\it Test set {\rm versus} Training set:}
\begin{equation}
Q_{pt} = U_p U_t^\top  = K_{pt} W W^\top K_{tt}^\top\;,
\end{equation}

{\it Test set {\rm versus} Test set:}
\begin{equation}
Q_{pp} = U_p U_p^\top  = K_{pt} W W^\top K_{pt}^\top\;.
\end{equation}

Here we want to find the $n \times m$ weight matrix $W$ such that $Q_{tt}$ fits $G_{tt}$ as much as possible.
If we set $A = W W^\top$, this problem can be replaced by finding $A$ which minimizes the difference between $G_{tt}$ and $Q_{tt}$.
It means that, this enables us to avoid considerable computational burden for computing $W$ itself, even if $m$ is infinite.
Therefore, we attempt to find $A (= W W^\top)$ which minimizes
\begin{equation}
L = \parallel G_{tt} - K_{tt} A K_{tt}^\top \parallel_F^2,
\end{equation}
where $\parallel \cdot \parallel_F$ indicates the Frobenius norm.
We can rewrite the above equation in the trace form as 
\begin{equation}
L = {\rm tr} \left\{ (G_{tt} - K_{tt} A K_{tt}^\top)(G_{tt} - K_{tt} A K_{tt}^\top)^\top \right\}.
\end{equation} 
The derivative of $L$ with respect to $A$ is obtained as
$$
\frac{1}{2} \frac{\partial L}{\partial A} = - K_{tt} G_{tt} K_{tt} + K_{tt}^2 A K_{tt}^2 ,
$$

From setting $\frac{\partial L}{\partial A} = 0$, the solution is analytically obtained by 
$$
A = W W^\top = K_{tt}^{-1} G_{tt} K_{tt}^{-1}.
$$
Then, the feature score matrix $U$ can be computed for training set and test set, respectively, as follows: 
\begin{equation}
U_t =  K_{tt} K_{tt}^{-1} G_{tt}^{1/2}, \quad
U_p =  K_{pt} K_{tt}^{-1} G_{tt}^{1/2}.
\end{equation}
Therefore, we can compute the feature-based similarity matrix $Q$ involving the test set as follows:

{\it Test set {\rm versus} Training set:}
\begin{equation}
Q_{pt} = U_p U_t^\top  = K_{pt} K_{tt}^{-1} G_{tt},
\end{equation}

{\it Test set {\rm versus} Test set:}
\begin{equation}
Q_{pp} = U_p U_p^\top  = K_{pt} K_{tt}^{-1} G_{tt} K_{tt}^{-1} K_{pt}^\top.
\end{equation}
By using the $Q_{pt}$ and $Q_{pp}$, we can predict the missing entries in the kernel matrix $G$, which correspond to $G_{pt}$ and $G_{pp}$.

\subsection{Penalized kernel matrix regression (PKMR)}
\label{sec:PKMR}

Here we consider introducing the idea of regularization in the KMR proposed in the previous section.
To do so, we attempt to find $A (= W W^\top)$ which minimizes the following penalized loss function:
\begin{equation}
L = \parallel G_{tt} - K_{tt} A K_{tt} \parallel_F^2 + \lambda PEN(A),
\end{equation}
where $\lambda$ is a regularization parameter and $PEN(A)$ is a penalty term for $A$ defined as follows. Each positive semidefinite matrix $A$ can be expanded as $ A = \sum_{i=1}^n \wb_i \wb_i^\top$, 
where the $(\wb_i)_{i=1,\ldots,n}$ form an orthogonal basis of eigenvectors. To each $\wb_i$ is associated a feature $u_i:\RR^d\rightarrow\RR$ by (\ref{eq:feature}), whose norm in the RKHS of $K$ is given by:
$$
\parallel u_i \parallel_{RKHS}^2 = \sum_{j,k=1}^n w_{i,j}w_{i,k}k(\xb_j,\xb_k) = {\rm tr}(\wb_i\wb_i^\top K) \;.
$$
To enforce regularity of the global mapping $\ub$, we therefore define the following penalty for $A$:
$$
PEN(A) = 2 \sum_{i=1}^n \parallel u_i \parallel_{RKHS}^2  
       = 2 \sum_{i=1}^n {\rm tr}(\wb_i\wb_i^\top K_{tt} )
       = 2 {\rm tr} (A K_{tt} ) \;.
$$

In this case, the optimization problem is reduced to finding $A$ which minimizes
\begin{equation}
L = {\rm tr} \left\{ (G_{tt} - K_{tt} A K_{tt}^\top)(G_{tt} - K_{tt} A K_{tt}^\top)^\top \right\}
 + 2 \lambda {\rm tr} \left\{ A K_{tt} \right\}.
\end{equation} 

The derivative of $L$ with respect to $A$ is 
$$
\frac{1}{2} \frac{\partial L}{\partial A} = - K_{tt} G_{tt} K_{tt} + K_{tt}^2 A K_{tt}^2 + \lambda K_{tt}.
$$
Therefore, the solution of the above penalized optimization problem is obtained by 
$$
A = K_{tt}^{-1} (G_{tt} - \lambda K_{tt}^{-1}) K_{tt}^{-1}.
$$
We note that the justification for the penalty used is only valid for positive semidefinite matrices, which will be obtained at least for small enough $\lambda$. Therefore, we can compute the feature-based similarity matrix $Q$ involving the test set as follows:

{\it Test set {\rm versus} Training set:}
\begin{equation}
Q_{pt} = U_p U_t^\top  = K_{pt} K_{tt}^{-1} (G_{tt} - \lambda K_{tt}^{-1}),
\end{equation}

{\it Test set {\rm versus} Test set:}
\begin{equation}
Q_{pp} = U_p U_p^\top  = K_{pt} K_{tt}^{-1} (G_{tt} - \lambda K_{tt}^{-1}) K_{tt}^{-1} K_{pt}^\top.
\end{equation}
By using the $Q_{pt}$ and $Q_{pp}$, we can predict the missing entries in the kernel matrix $G$, which correspond to $G_{pt}$ and $G_{pp}$.

\section{Relationship with the {\it em}-algorithm}
\label{sec:em}

For the kernel matrix completion problem, the use of the $em$ algorithm based on information geometry has been proposed [7].
There the kernel matrix completion problem is defined as finding missing entries that minimize the Kullback-Leibler divergence between the resulting completed matrix and a spectral variant of the full matrix.

It is interesting to observe that the final algorithms between $em$ and KMR are very similar. The $em$ algorithm results in the following equations for estimating the incomplete parts $G_{pt}$ and $G_{pp}$ in $G$:

{\it Test set {\rm versus} Training set:}
\begin{equation}
Q_{pt} = K_{pt} K_{tt}^{-1} G_{tt},
\end{equation}

{\it Test set {\rm versus} Test set:}
\begin{equation}
Q_{pp} =
K_{pp} + 
K_{pt} K_{tt}^{-1} K_{pt}^\top +
K_{pt} K_{tt}^{-1} G_{tt} K_{tt}^{-1} K_{pt}^\top.
\end{equation}

We note that the $Q_{pt}$ of the {\it em}-algorithm is equivalent to that of the kernel matrix regression.
On the other hand, the $Q_{pp}$ of the {\it em}-algorithm is {\it not} equivalent to that of the kernel matrix regression.
It differs by $K_{pp} + K_{pt} K_{tt}^{-1} K_{pt}^\top$.
This stems from the difference of the geometry space between the two methods.
The {\it em}-algorithm is based on the information geometry, while the proposed KMR is based on the Euclidean geometry.

\section{Experiments}
\label{sec:experiments}

In this section we report an empirical comparison of different methods: 1) direct method, 2) kCCA method, 3) {\it em}-method, 4) KMR method, 5) PKMR method applied to the problem of gene network inference and protein 3D structure prediction.

\subsection{Estimation of missing edges in the metabolic gene network}

The metabolic gene network is an important biological network.
However, most parts of the metabolic gene network remain unknown, and many enzyme genes are still missing in our current knowledge. Determining new enzyme genes and their position in the metabolic network is an expensive and painstaking process that requires many wet experiments.
On the contrary, we can easily obtain various genome-wide genomic datasets representing gene/protein information, such as gene expression data, protein localization data, and phylogenetic profiles.
We therefore attempt to predict missing edges in the metabolic gene network by using such genomic data.

We gathered a kernel matrix of the genomic data (consisting of 769 genes) by combining three kernel matrices obtained from three datasets: gene expression data, protein localization data, and phylogenetic profiles, and we regard this matrix as an explanatory kernel matrix $K$.
We used the same datasets and corresponding kernels as those used by [4]. 
We obtained a kernel matrix for the gene network from the graph information of the gene network by using the diffusion kernel [8] with parameter $\sigma=1$, and we regard this matrix as a response kernel $G$.
The kernel matrix $K$ is invertible in this case.
The kernel similarity values in $G$ (transformed by the diffusion kernel from the graph of the gene network) are expected to represent the intensity of graphical association between genes, which can be considered as a possibility of the existence of the edge.
Therefore, if the gene pairs sharing similarities higher than a threshold, they are predicted to interact with each other.

To compare the performance between different methods, we applied the direct approach, the kCCA, the {\it em}-algorithm, the KMR, and the penalized KMR (PKMR) to the gene network prediction.
We tested their performance by cross-validation.
In each cross-validation iteration, we randomly split the genes in the gold standard data into training set and test set. We learned the model based on the training dataset only and we applied the model to the test set in order to predict the missing edges involving the test set on the metabolic network.
We are also interested in the effect of the rate of the test samples against the training samples, so we carried out the same experiment with different percentages of the test samples in the data splitting process in each cross-validation iteration.

As a measure of the performance, we used the AUC score (area under the ROC curve), because the performance depends on the threshold given in advance.
The ROC curve is defined as a function of the true positive rates against the false positive rates based on several threshold values.
"True positive" means that the predicted gene-pairs are actually present in the gold standard network, while "false positive" means that the predicted gene-pairs are absent in the gold standard network.
In the case of the kCCA, we set the regularization parameters $\lambda_x$ and $\lambda_y$ as 0.1 and 0.1, and we used 30 features, as suggested by [4].
In the case of the PKMR, the regularization parameter $\lambda$ is optimized by applying the internal cross-validation within the training set with the AUC score as a target criterion, which provides us with $\lambda=0.1$.

All the result of the experiments are summarized in Tables 1 and 2.
Table 1 shows the result in estimating $Q_{pt}$, while Table 2 shows the result in estimating $Q_{pp}$.
In each table, the rows correspond to the percentage of the training samples against the test samples and the columns correspond to the methods.
It appears that the direct approach performs significantly worse than the other supervised learning based methods.
It seems that, the kCCA performs worse than the {\it em}-algorithm, the KMR and PKMR in estimating $Q_{pt}$, while the kCCA performs better than the {\it em}-algorithm and the KMR, and at competitive level with the PKMR in estimating $Q_{pp}$.
Focusing on the comparison of the performance between the {\it em}-algorithm and the KMR, both the {\it em}-algorithm and the KMR show the same performance in estimating $Q_{pt}$, as expected from the mathematical relationship between the {\it em}-algorithm and the KMR.
On the other hand, the {\it em}-algorithm and the KMR behave differently in estimating $Q_{pp}$, but their performances are at competitive level.
The penalized KMR (PKMR) slightly outperforms the other methods in estimating both $Q_{pt}$ and $Q_{pp}$, suggesting that the introduction of regularization can be meaningful in this context.

\begin{table}[t]
\caption{Comparison of AUC scores with varying rates of training set: $Q_{tp}$ (training versus test).}
\label{auc_tp}
\vskip 0.15in
\begin{center}
\begin{small}
\begin{sc}
\begin{tabular}{lccccr}
\hline
Rate & direct & kCCA & em & KMR & PKMR \\
\hline
90 \% & 0.598 & 0.840 & 0.889 & 0.889 & {\bf 0.892} \\
80 \% & 0.570 & 0.824 & 0.844 & 0.844 & {\bf 0.848} \\
70 \% & 0.580 & 0.783 & 0.805 & 0.805 & {\bf 0.814} \\
60 \% & 0.575 & 0.780 & 0.786 & 0.786 & {\bf 0.801} \\
50 \% & 0.579 & 0.772 & {\bf 0.783} & {\bf 0.783} & 0.772 \\
40 \% & 0.569 & 0.714 & {\bf 0.760} & {\bf 0.760} & 0.758 \\
30 \% & 0.571 & 0.682 & 0.732 & 0.732 & {\bf 0.738} \\
20 \% & 0.565 & 0.633 & 0.674 & 0.674 & {\bf 0.676} \\
10 \% &  0.593 & 0.669 & 0.672 & 0.672 & {\bf 0.676} \\
\hline
\end{tabular}
\end{sc}
\end{small}
\end{center}
\vskip -0.1in
\end{table}

\begin{table}[t]
\caption{Comparison of AUC scores with varying rates of training set: $Q_{pp}$ (test versus test).}
\label{auc_pp}
\vskip 0.15in
\begin{center}
\begin{small}
\begin{sc}
\begin{tabular}{lcccccr}
\hline
Rate & direct & kCCA & em & KMR & PKMR \\
\hline
90 \% & 0.531 & 0.785 & 0.766 & 0.774 & {\bf 0.787} \\
80 \% & 0.593 & 0.727 & 0.724 & 0.723 & {\bf 0.743} \\
70 \% & 0.602 & 0.680 & 0.686 & 0.700 & {\bf 0.703} \\
60 \% & 0.558 & 0.673 & {\bf 0.683} & 0.678 & 0.675 \\
50 \% & 0.581 & 0.661 & 0.644 & 0.651 & {\bf 0.662} \\
40 \% & 0.569 & {\bf 0.646} & 0.635 & 0.635 & 0.642 \\
30 \% & 0.583 & 0.610 & 0.621 & 0.627 & {\bf 0.637} \\
20 \% & 0.579 & {\bf 0.587} & {\bf 0.587} & 0.567 & 0.576 \\
10 \% & 0.568 & 0.591 & 0.585 & 0.573 & {\bf 0.589} \\
\hline
\end{tabular}
\end{sc}
\end{small}
\end{center}
\vskip -0.1in
\end{table}




\subsection{Prediction of protein 3D structures from their sequences}

Protein 3D structures are strongly associated with evolutionary history and biological functions, compared with protein sequences. 
Here we attempt to classify proteins into superfamilies based on the structure information.
However, the number of proteins whose structures are determined is limited even nowadays and the structure information of most proteins is almost unknown.
Therefore, we performed protein classification by predicting missing similarity elements of protein 3D structures.

The sequence similarities are obtained by marginalized kernel [9] and the 3D structural similarities are obtained by the result of MATRAS [10].
We used the same datasets and corresponding kernel matrices used in previous work [5], where $K$ corresponds to the similarity matrix for protein sequences and $G$ corresponds to the similarity matrix for protein 3D structures.
The kernel matrix $K$ is invertible in this case.
We applied the support vector machine (SVM) to the dataset of TIM beta/alpha-barrel protein fold (18 classes, 90 proteins), and conducted one-versus-other supervised classifications for class 1 and class 3, respectively.
The reason why we selected the above two classes is that they have more than 10 members in each class. 
The leave-one-out cross-validation is conducted and the performance is evaluated by using sensitivity and specificity, where sensitivity is defined as $\#TP / (\#FN + \#TP)$ and specificity is defined as $\#TP / (\#FP + \#TP)$, respectively. 
In the case of PKMR, the regularization parameter $\lambda$ is optimized by applying the internal cross-validation within the training set, and set to be 0.1. 

\begin{table}[h]
\caption{Comparison of sensitivities (left) and specificities (right): Class 1.}
\begin{center}
\begin{tabular}{cc}
\begin{minipage}[t]{7.2cm}\begin{tabular}{lccccr}
\hline
Rate & direct & kCCA & em & KMR & PKMR \\
\hline
 0\% & 0.47 & - & - & - & - \\
20\% & - & 0.82 & 0.76 & 0.76 & {\bf 0.88} \\
40\% & - & 0.70 & 0.70 & 0.70 & {\bf 0.76} \\
60\% & - & 0.70 & 0.88 & 0.88 & {\bf 0.94} \\
80\% & - & 0.76 & {\bf 0.88} & {\bf 0.88} & {\bf 0.88} \\
100\% &  0.94 & 0.94 & 0.94 & 0.94 & 0.94 \\
\hline
\end{tabular}
\end{minipage}
&
\begin{minipage}[t]{7.2cm}
\begin{tabular}{lccccr}
\hline
Rate & direct & kCCA & em & KMR & PKMR \\
\hline
 0\% & 0.66 & - & - & - & - \\
20\% & - & {\bf 0.45} & 0.43 & 0.41 & 0.42 \\
40\% & - & 0.36 & {\bf 0.4}  & 0.36 & 0.38 \\
60\% & - & 0.5 & {\bf 0.53} & 0.48 & 0.51 \\
80\% & - & 0.81 & {\bf 0.93} & {\bf 0.93} & {\bf 0.93} \\
100\% &  1.0 & 1.0 & 1.0 & 1.0 & 1.0 \\
\hline
\end{tabular}
\end{minipage}
\end{tabular}
\end{center}
\end{table}

\begin{table}[h]
\caption{Comparison of sensitivities (left) and specificities (right): Class 3.}
\begin{center}
\begin{tabular}{cc}
\begin{minipage}[t]{7.2cm}
\begin{tabular}{lccccr}
\hline
Rate & direct & kCCA & em & KMR & PKMR \\
\hline
 0\% & 0.58 & - & - & - & - \\
20\% & - & 0.64 & 0.76 & 0.76 & {\bf 0.82} \\
40\% & - & 0.64 & 0.82 & {\bf 0.88} & {\bf 0.88} \\
60\% & - & 0.76 & 0.82 & 0.82 & {\bf 0.88} \\
80\% & - & 0.76 & 0.94 & 0.94 & {\bf 1.0} \\
100\% &  1.0 & 1.0 & 1.0 & 1.0 & 1.0 \\
\hline
\end{tabular}
\end{minipage}
&
\begin{minipage}[t]{7.2cm}
\begin{tabular}{lccccr}
\hline
Rate & direct & kCCA & em & KMR & PKMR \\
\hline
 0\% & 0.62 & - & - & - & - \\
20\% & - & 0.40 & {\bf 0.44}   & 0.34 & 0.35 \\
40\% & - & 0.39 & {\bf 0.53}  & 0.45 & 0.42 \\
60\% & - & {\bf 0.56} & 0.51  & 0.51 & 0.53 \\
80\% & - & 0.68 & {\bf 0.94}  & {\bf 0.94} & 0.85 \\
100\% &  0.94 & 0.94 & 0.94 & 0.94 & 0.94 \\
\hline
\end{tabular}
\end{minipage}
\end{tabular}
\end{center}
\end{table}

Table 3 and Table 4 show the results of computing sensitivities and specificities depending on the rate of training set for class 1 and class 3, respectively, in the leave-one-out cross-validation experiments.
The direct method with $0 \%$ means that we use sequence information only, while the direct method with $100 \%$ means that we use structure information only.
Looking at the tables, the PKMR seems to outperform the other methods in this context especially from the viewpoint of sensitivity.
In contrast, the performance seems to be at competitive level across different methods from the viewpoint of specificity.

\section{Discussions and conclusions}
\label{sec:discussions}

In this paper we addressed the problem of filling missing entries in a kernel Gram matrix, given a related full kernel Gram matrix.
We attacked this kernel matrix completion problem from the viewpoint of regression.
We proposed the kernel matrix regression (KMR) based on the underlying features in the reproducing kernel Hilbert space by modifying the idea of kCCA.
Through the development of the KMR, we also clarified the mathematical relationship between the kCCA, the {\it em}-algorithm, and the proposed methods in the context of the kernel matrix completion problem.
In the experiment on gene network inference and protein 3D prediction, we confirmed that the performance of the KMR is competitive with that of other methods, and we showed that the penalized version of the KMR works the best when an appropriate regularization parameter is chosen.

\subsubsection*{References}

\normalsize{
%
%

\noindent [1] Schoelkopf, B. \& Smola, A.J. (2002) {\it Learning with Kernels: Support Vector Machines, Regularization, Optimization, and Beyond.}
Cambridge, MA: MIT Press.

\noindent [2] Schoelkopf, B., Tsuda, K. \& Vert, J.-P. (2004) {\it Kernel Methods in Computational Biology.}
Cambridge, MA: MIT Press.

\noindent [3] Vert, J.-P. \& Kanehisa, M. (2003) Graph-driven features extraction from microarray data using diffusion kernels and kernel CCA. {\it Advances in Neural Information Processing
Systems 15}, pp. 1449-1456. Cambridge, MA: MIT Press.

\noindent [4] Yamanishi, Y., Vert, J.-P. \& Kanehisa, M. (2004) Protein network inference from multiple genomic data: a supervised approach. {\it Bioinformatics}, {\bf 20}:i363-i370.

\noindent [5] Kim, T., Kato, T. \& Tsuda, K. (2004) Protein classification via Kernel Matrix Completion. In Schoelkopf, B., Tsuda, K. \& Vert, J.-P. (eds.), {\it Kernel Methods in Computational Biology.}, pp. 261-276. Cambridge, MA: MIT Press.

\noindent [6] Bach, F., \& Jordan, M. (2002) Kernel Independent Component Analysis. {\it Journal of Machine Learning Research}
{\bf 3}:1-48.

\noindent [7] Tsuda, K., Akaho, S. \& Asai, K. (2003) The em Algorithm for Kernel Matrix Completion with Auxiliary Data. {\it Journal of Machine Learning Research}
{\bf 4}:67-81.

\noindent [8] Kondor, R.I. \& Lafferty, J. (2002) Diffusion Kernels on Graphs and Other Discrete Input. {\it ICML2002}.

\noindent [9] Tsuda, K., Kim, T. \& Asai, K. (2002) Marginalized kernels for biological sequences. {\it Bioinformatics}
{\bf 18}:S268-S275.

\noindent [10] Kawabata, T. \& Nishikawa, K. (2000) Protein tertiary structure comparison using the Markov transition model of evolution. {\it Proteins}
{\bf 41}:108-122.

}

\end{document}